\documentclass[sigconf]{acmart}

\usepackage[utf8]{inputenc}
\usepackage{hyperref}
\usepackage{url}
\usepackage{booktabs}
\usepackage{amsfonts}
\usepackage{microtype}
\usepackage{amsmath,amssymb,amsthm,bm}
\usepackage{subfigure}

\usepackage{IEEEtrantools}
\DeclareMathOperator*{\argmax}{arg\,max}

\AtBeginDocument{%
  \providecommand\BibTeX{{%
    \normalfont B\kern-0.5em{\scshape i\kern-0.25em b}\kern-0.8em\TeX}}}

\setcopyright{acmcopyright}
\copyrightyear{2019}
\acmYear{2019}
\acmDOI{10.1145/3356464.3357707}

\acmConference[DAI '19]{First International Conference on Distributed Artificial Intelligence}{October 13--15, 2019}{Beijing, China}
\acmPrice{15.00}
\acmISBN{978-1-4503-7656-3/19/10}

\begin{document}

\title{Factorized Q-Learning for Large-Scale Multi-Agent Systems}

\author{Ming Zhou$^{1}$, Yong Chen$^2$, Ying Wen$^3$, Yaodong Yang$^3$, Yufeng Su$^1$,\\ Weinan Zhang$^1$, Dell Zhang$^4$, Jun Wang$^3$}
\affiliation{
	$^1$Shanghai Jiao Tong University, $^2$Beihang University, $^3$University College London, $^4$Birkbeck, University of London
}
\email{
	{mingak, suyufeng1997, wnzhang}@sjtu.edu.cn,
	chenyong@nlsde.buaa.edu.cn,
	{yin.wen, yaodong.yang, jun.wang}@cs.ucl.ak.uk,
	dell.z@ieee.org
}








\renewcommand{\shortauthors}{Ming Zhou, et al.}

\newtheorem{assumption}{\textbf{Assumption}}

\begin{abstract}
 Deep Q-learning has achieved significant success in single-agent decision making tasks.
However, it is challenging to extend Q-learning to large-scale multi-agent scenarios, due to the explosion of action space resulting from the complex dynamics between the environment and the agents.
In this paper, we propose to make the computation of multi-agent Q-learning tractable by treating the Q-function (w.r.t. state and joint-action) as a high-order high-dimensional tensor and then approximate it with factorized pairwise interactions.
Furthermore, we utilize a composite deep neural network architecture for computing the factorized Q-function, share the model parameters among all the agents within the same group, and estimate the agents' optimal joint actions through a coordinate descent type algorithm.
All these simplifications greatly reduce the model complexity and accelerate the learning process.
Extensive experiments on two different multi-agent problems demonstrate the performance gain of our proposed approach in comparison with strong baselines, particularly when there are a large number of agents.
\end{abstract}




\maketitle

\section{Introduction}
Multi-Agent Reinforcement Learning (MARL) studies a group of autonomous agents in a shared environment from which they learn what to do according to the received reward signals while interacting with each other.
  For many real-world applications, it is appealing to employ multiple agents because they could accomplish tasks that a standalone agent could not do or would do in a costly manner.
  
  The great obstacle for applying single-agent reinforcement learning algorithms such as Q-learning \cite{Christopher-PhD-Thesis-1989} directly to the multi-agent setting is that with the presence of other agents taking actions, the environment for each individual agent can no longer be regarded as stationary.
  To address the difficult decision problems arising from MARL, researchers have tried to borrow techniques from game theory, in particular, the framework of stochastic games \cite{Littman-MarkovGames-ICML-1994}.
  However, such algorithms are computationally expensive and therefore only able to deal with a few agents.
  
  In this paper, we aim to make Q-learning for MARL scalable to a large number of agents.
  Inspired by the Factorization Machines \cite{Steffen-FM-TIST-2012,Steffen-PITensor-WSDM-2010} widely used in recommender systems, we model the complex relationship between the environment and the agents as a high-order high-dimensional tensor and then approximate it through factorization.
  Specifically, the multi-agent Q-function (w.r.t.  state and joint-actions) is decomposed into independent components plus pairwise interactions (between any two agents). 
  As indicated in \cite{Lawrence-Strategic-Interaction-Games-1993}, focusing on pairwise interactions could greatly reduce the complexity of a multi-agent system while maintaining the essence of the multidimensional complex relationship among different agents.
  Moreover, such a factorized Q-function is going to be shared among different agents within the same group (or the entire system when there is no grouping of agents), which further cuts down the complexity of the multi-agent system and also helps to speed up the deep learning process.
  It is also worth noting that the agents' last actions are leveraged to estimate their current strategies in the optimization algorithm, which effectively mitigates the combinatorial explosion of joint actions.
  In summary, we propose a computationally efficient Q-function approximation for MARL named ``Factorized Q-learning (FQL)'' which is capable of handling large-scale multi-agent systems.

\section{Related Work}

\subsection{Single-Agent Q-Learning}
  Q-Learning \cite{Christopher-PhD-Thesis-1989,Christopher-Technical-QLearning-ML-1992,Francisco-Convergence-Report-2001} is a model-free off-policy reinforcement learning method that estimates the long-term expected return of executing an action $a$ from a given state $s$. 
  The estimated returns, known as Q-values, can be learned iteratively by updating the current Q-value estimate towards the observed reward $r_t$ plus the maximum possible Q-value over all actions $a$ in the next state $s_{t+1}$:
  
  \begin{equation}
    Q(s_t,a_t) \gets (1-\alpha) Q(s_t,a_t) + \alpha  \left( r_t + \gamma \cdot \max_a Q(s_{t+1},a) \right) \ , \nonumber
  \end{equation}
  where $\gamma \in [0,1)$ is the discount factor and $\alpha \in (0,1]$ is the learning rate.
  
  For challenging domains like Atari games, there are too many states to allow us to maintain all the Q-values in a table, so a model is needed instead for the computation of the Q-function.
  The state of the art solution is the Deep Q-Network (DQN) algorithm \cite{Volodymyr-DQN-Nature-2015} which approximates as well as generalizes the relationship between states (inputs) and actions (outputs) with a deep neural network $Q(s,a;\theta)$ parameterized by $\theta$.
  The network parameters are learned via back propagation to minimize a differentiable loss function --- the squared \emph{temporal difference} error 
  
  \begin{equation}
    L(\theta) = \mathbb{E}_{(s_t,a_t,r_t,s_{t+1}) \sim Unif(\mathcal{D})} \left[ {\left( {Y_t - Q(s_t,a_t;\theta)} \right)}^2 \right] \;
  \end{equation}
  with
  \begin{equation}
    Y_t = r_t + \gamma \cdot \max_a Q(s_{t+1},a;\tilde{\theta}) \ ,
  \end{equation}
  where $(s_t,a_t,r_t,s_{t+1})$ are the past experiences recorded in a ``replay memory'' $\mathcal{D}$ and then sampled uniformly from $\mathcal{D}$ to train the network in a supervised manner, while $\tilde{\theta}$ represents the parameters of a ``target network'' $\widetilde{Q}$ that are periodically copied from the online neetwork $Q$ and kept constant for a number of iterations in order to make the DQN training stable.
  
  In noisy environments, Q-learning often overestimates the action values, which may slow the learning down \cite{Hado-DoubleQLearning-NIPS-2010}. 
  This problem could be alleviated by the ``double DQN'' trick \cite{Hado-DDQN-AAAI-2016} that uses the online network $Q$ to select the next action but the target network $\widetilde{Q}$ to estimate its value:
  
  \begin{equation}
    Y_t = r_t + \gamma \cdot Q(s_{t+1},\argmax_a{Q(s_{t+1},a;\theta)};\tilde{\theta}) \ . 
    \label{eq:double_DQN}
  \end{equation}
  
\subsection{Multi-Agent Q-Learning}
  Generally speaking, the MARL algorithms that try to solve the multi-agent stochastic games can be divided into two paradigms: \emph{equilibrium learning} and \emph{best-response learning}.
  
  In the equilibrium learning paradigm, the agents try to learn policies which form a Nash equilibrium \cite{Michael-Team-Q-CognitiveSystems-2001}.
  Specifically, each agent attempts to get at least the amount of payoff indicated by a Nash equilibrium, i.e., the lower-bound of performance, regardless of the policies being played by the other players.
  Since it is usually difficult to find such equilibrium, existing algorithms focus on a small class of stochastic games, e.g., zero-sum games or two-person general-sum games.
  For example, Nash-Q \cite{Junling-NashQLearning-JMLR-2003} and Friend-or-Foe \cite{Michael-FFQ-ICML-2001} extend the classic Q-learning \cite{Christopher-Technical-QLearning-ML-1992} by encoding the interactions between environment and agents in a so-called Nash Q function.
  It is proved that Nash Q learning converges to the optimal policy under some restrictive assumptions.
  This is also the case for the recently emerged mean-field Q-learning (MF-Q) algorithm \cite{Yang2018} which equips each agent with one Q-function and approximates it by the average effects in the agent's neighborhood.
  However, such algorithms are not practical in a complex environment with a large number of agents because of the expensive computation required to estimate other agents' policies at each state and find the equilibrium.
  Besides, when there are many agents, the estimated policies of different agents might not belong to the same Nash equilibrium, thus the convergence will become invalid \cite{Gerald-Hyper-Q-NIPS-2003}.
  
  In the best-response learning paradigm, each agent just tries to learn a policy that is optimal with respect to the joint policy of the other players \cite{Claus1998,Uther1997}.
  On one hand, such methods are not assured of the lower-bound of performance, especially when the other agents do not have stationary policies.
  On the other hand, it is possible for an agent to take advantage of the fact that the policies being played by the other players may not be their best responses and thus obtain more reward than that guaranteed by the equilibrium.
  The simplest algorithms in this category back off to the single-agent case and just conduct independent Q-learning (IQL) in which each agent independently learns its own Q-function by treating the other agents as part of the environment without considering the interactions among different agents \cite{Ming-MARL-ICML-1993,Ardi-MultiagentDQN-PLOSONE-2017}. 
  RIAL (Reinforced Inter-Agent Learning) \cite{Jakob-RIAL-NIPS-2016} combines the idea of IQL with DRQN \cite{Guillaume-DRQN-FPS-AAAI-2017} to learn communication protocols in a cooperative multi-agent environment.
  Similarly, multi-agent DQN (MA-DQN) \cite{Ardi-MultiagentDQN-PLOSONE-2017} carries out IQL with an autonomous DQN for each agent to investigate the interaction between two agents in the video game Pong.
  Although the IQL style algorithms are computationally efficient and therefore can accommodate a large number of agents, they are often sub-optimal because, as we have mentioned above, the environment would be non-stationary from each agent's point of view.
  Noticeably, the additional knowledge about the other agents should be beneficial to the effectiveness of learning, and sharing policies or episodes among the agents could speed up the learning process \cite{Ming-MARL-ICML-1993}.
  Value-Decomposition Networks (VDN) \cite{Peter-VDNs-CoRR-2017} goes a little bit beyond IQL by summing over all the independent Q-functions for cooperative tasks, but the complex interactions in MARL are unlikely to be captured by simplistic linear summations.
  Monotonic Value Function Factorization (QMIX) \cite{Tabish-QMIX-CoRR-2018} mixes the per-agent action-value Q-functions into a rich joint action-value function which provides extra state information for learning.
  However, the QMIX architecture will become more and more complicated and difficult to compute as the number of agents increases.
  In contrast, the complexity of our proposed method depends not on the number of agents but on the number of agent-groups which usually remains to be small even for large-scale MARL problems.
  
\section{Multi-Agent Factorized Q-Learning}
  Here we extend the Deep Q-Network (DQN) \cite{Volodymyr-DQN-Nature-2015} to multi-agent environments with an approximate Q-function based on Factorization Machines \cite{Steffen-FM-TIST-2012,Steffen-PITensor-WSDM-2010}. 
  Specifically, we first reformulate the multi-agent joint-action Q-function in a factorized form, then present the optimization algorithm for learning such factorized Q-functions through deep neural networks, and finally provide an analysis of this algorithm's computational complexity.
  
\subsection{Multi-Agent Q-Function Approximation}
  DQN combines Q-learning and deep neural networks to conduct single-agent reinforcement learning and has achieved phenomenal success in playing computer games etc.
  How to extend the successful DQN technique to multi-agent systems is an important research problem with numerous potential applications. 
  The significant difference between single-agent and multi-agent reinforcement learning is that the former only needs to consider individual actions of one agent whereas the latter should take the complex interactions among multiple agents into account to optimize the \emph{joint actions}. 
  For an $N$-agent system, the multi-agent Q-function for each agent should be $Q(s,a^1,a^2,\cdots,a^N)$ with $s \in \mathcal{S} = {\mathcal{S}^1} \times {\mathcal{S}^2} \times \cdots \times {\mathcal{S}^N}$ representing the overall state and $(a^1,a^2,\cdots,a^N) \in \mathcal{A} = {\mathcal{A}^1} \times {\mathcal{A}^2} \times \cdots \times {\mathcal{A}^N}$ representing the join action, where $\mathcal{S}^i$ and $\mathcal{A}^i$ are the $i$-th agent's individual state space and individual action space respectively.
  At time-step $t$, agent $i$'s own state is $s^i_t$ and its own action is $a^i_t$, while all agents' overall state is $s_t$ and their joint action is $a_t = (a^1_t,a^2_t,\cdots,a^N_t)$.

  To make the Q-function computation scalable to a large number of agents, we make two fundamental assumptions. 

  \paragraph{Low Intrinsic Dimensionality.} In the reinforcement learning literature, the low-dimensionality (low-rank) assumption is widely regarded valid for single-agent, and researchers have utilized various approximation schemes to compactly represent the Q-function, e.g., Radial Basis Function (RBF), Cerebellar Model Articulation Controller (CMAC) \cite{sutton1998reinforcement}, Product of Experts (PoE) \cite{sallans2004reinforcement}, and Robust Principle Component Analysis (RPCA) \cite{ong2015value}. It is reasonable to believe that the low-dimensionality assumption holds for multi-agent Q-learning as well. Considering the particular structure of the multi-agent Q-function which involves the complex relationship among many agents, we hereby propose to find its low-dimensional approximation by borrowing the idea from the Factorization Machines \cite{Steffen-FM-TIST-2012,Steffen-PITensor-WSDM-2010} which captures the complex relationship among many users (items) with only independent components plus pairwise interactions.
    
  \paragraph{Parameter Sharing.} We assume all the agents in the same group are homogeneous and let them share the same model (Q-function) rather than maintaining a separate model for each individual agent, which would obviously reduce the computational complexity a lot.

  Under the above two assumptions, we are able to greatly simplify the multi-agent Q-function as follows.
  
\begin{IEEEeqnarray}{rCl}
\IEEEeqnarraymulticol{3}{l}{
  Q^i(s,a^1,a^2,\ldots,a^N;\Theta)
} \nonumber \\
& \equiv &
  Q^i(s,a^i,a^{-i};\Theta)  \label{eq:step0}  \\
& \approx &
  Q^i(s,{a^i};\theta) + \lambda^o \cdot \sum_{j \in -i} V^i{{(s,{a^i};\beta_1)}^T} U^i(s,{a^j};\beta_2) \label{eq:step1} \\
& \approx &
  Q(s^i,a^i;\theta) + \lambda^o \cdot \sum_{j \in -i} V{(s^i,a^i;\beta_1)^T} U(s^j,a^j;\beta_2) \label{eq:step2} \\
& = &
  Q(s^i,a^i;\theta) + \lambda^o \cdot V{(s^i,a^i;\beta_1)^T} \sum_{j \in -i} U(s^j,a^j;\beta_2) \label{eq:step3} \\
& =  &
  Q(s^i,a^i;\theta) + \lambda \cdot V{(s^i,a^i;\beta_1)^T} \frac{\sum_{j \in -i} U(s^j,a^j;\beta_2)}{N-1} \IEEEeqnarraynumspace \label{eq:step4} \\
& =  &
  Q(s^i,a^i;\theta) + \lambda \cdot V{(s^i,a^i;\beta_1)^T} \overline{U}(s^{-i},{a^{-i}};\beta_2) \ , \label{eq:FQL}
\end{IEEEeqnarray}
where $-i$ is an index set ranging from $1$ to $N$ with $i$ removed, the learnable parameters $\Theta = \{\theta,\beta_1,\beta_2\}$,
the hyper-parameter $\lambda = \lambda^o (N-1)$, and $\overline{U}({s^{-i}},{a^{-i}};\beta_2) = \sum_{j \in -i} U(s^j,a^j;\beta_2)/(N-1)$.

The approximation made from Eq.~\eqref{eq:step0} to Eq.~\eqref{eq:step1} is of course based on the previous assumptions.
Eq.~\eqref{eq:step1} contains two terms which correspond to the independent component and the pairwise interactions respectively, while the high-order interactions (among three or more agents) have been ignored. 
The approximation made from Eq.~\eqref{eq:step1} to Eq.~\eqref{eq:step2} rests on the assumption that agents are homogeneous, so agent $i$'s individual model ($Q^i$, $V^i$, and $U^i$) is replaced by the shared model ($Q$, $V$, and $U$).
The remaining derivations from Eq.~\eqref{eq:step2} to Eq.~\eqref{eq:FQL} simplify the mathematical expression step by step.
In the end, the other agents' overall influence on agent $i$ is summarized into a compact form ${\overline{U}(s^{-i},{a^{-i}};\beta_2)}$.

Note that here we have used two separate vectors to represent agent $i$ in its pairwise interactions with other agents, the $i$-th column vector of $V$ and the $i$-th column vector of $U$, to facilitate the learning of factorization. 
This is in the same spirit of \texttt{word2vec} \cite{mikolov2013efficient} which learns two separate embeddings, one target embedding and one context embedding, for each word.

We refer to the final formula Eq.~\eqref{eq:FQL} as the factorized Q-function for agent $i$, and name our proposed approach to MARL ``Factorized Q-learning (FQL)''. 
Since the factorized Q-function for each agent requires the knowledge of the other agents' current states and their last actions for both training and execution, our proposed FQL technique mainly addresses the MARL problems with a central controller that communicates the global information to all the agents. 
Nevertheless, at any particular moment each agent is not supposed to know the other agents' current action choices, which makes our setting much more realistic than a completely centralized one \cite{Jelle-Sparse-Cooperative-Q-ICML-2004}.
The requirement of global knowledge will be further relaxed to the requirement of local neighborhood knowledge in the last game of our experiments. 

\subsection{Neural Architecture}

\begin{figure}[tb!]
	\centering
	\includegraphics[width=.9\columnwidth]{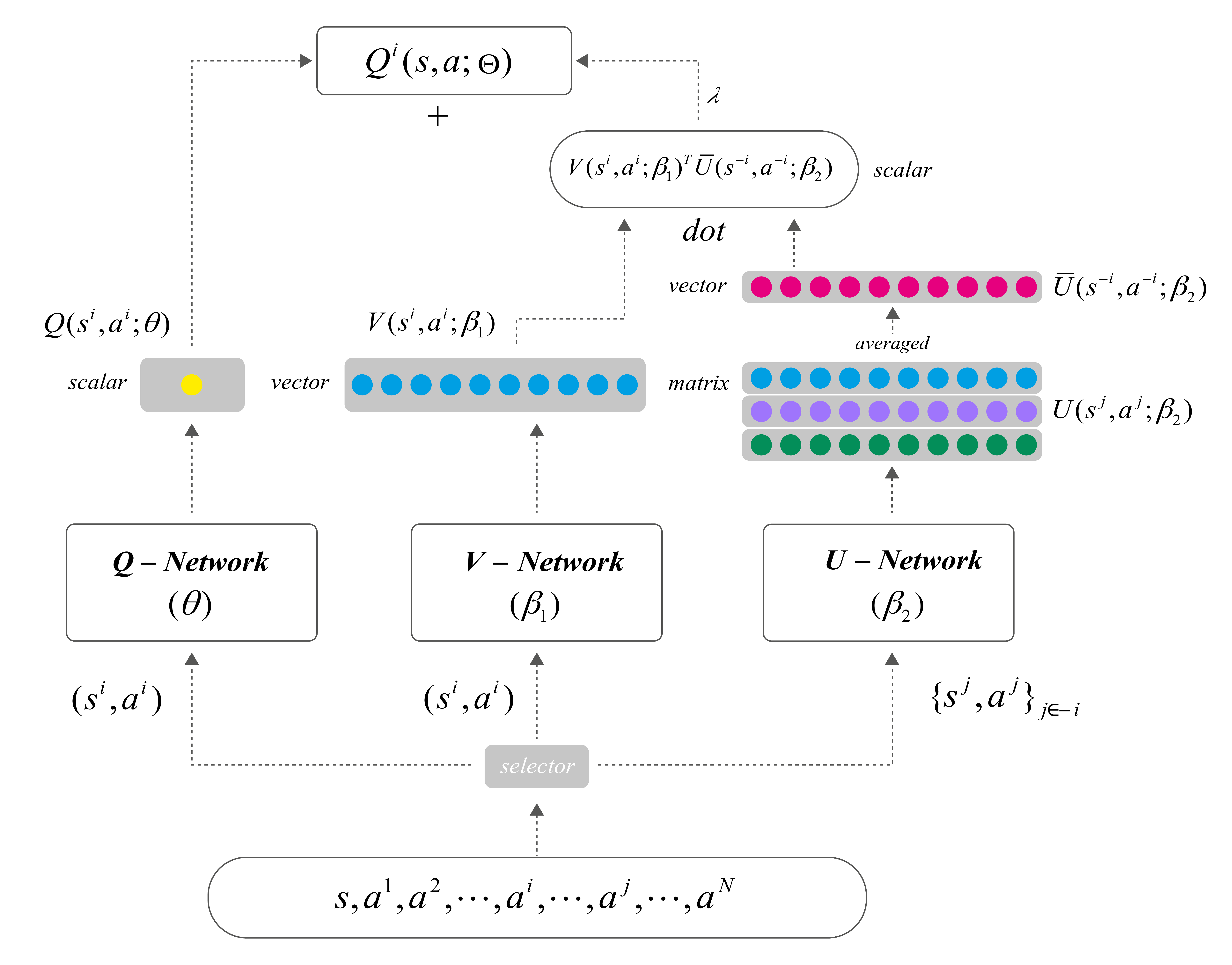}
	\caption{The composite deep neural network architecture for multi-agent Factorized Q-Learning (FQL).}
	\label{fig_model_structure}
\end{figure}

Similar to the DQN algorithm that we have described before, our proposed FQL approach employs a deep neural network to fit the Q-function for MARL.
The difference is that the FQL network is a composite one consisting of three sub-networks: $Q$-network, $V$-network and $U$-network corresponding to $Q(s^i,a^i;\theta)$, $V{(s^i,a^i;\beta_1)^T}$ and $\overline{U}(s^{-i},{a^{-i}};\beta_2)$ respectively in Eq.~\eqref{eq:FQL}.
The architecture of such a composite network is depicted in Fig~\ref{fig_model_structure}. 
The structural decomposition of the FQL network into three sub-networks makes its learning and inference much easier than a single big DQN network, in the MARL context.

\subsection{Optimization Algorithm}

Theoretically, for each agent $i$ at non-terminal time-step $t$, the target value $Y_t^i$ should be estimated as

\begin{equation}
\label{target_value_estimation_ideal}
Y_t^i = r_t^i + \gamma \cdot \widetilde{Q}^i({s_{t+1}},\argmax_{a^i,a^{-i}}{Q^i({s_{t+1}},\underbrace{a^i,a^{-i}}_\text{joint action};\Theta)};\widetilde{\Theta}) \ ,
\end{equation}
where ${\widetilde Q}^i$ is the target network, and $a^i$ and $a^{-i}$ represent the available actions for agents given the state $s_{t+1}$.
However, it is infeasible to directly search for the optimal joint action for state $s_{t+1}$ when there are many agents, as the size of the joint action space grows exponentially with $N$, the number of the agents in the system.

In order to make the computation tractable, we get the idea from \emph{coordinate descent} that the optimization of a multivariate function can be achieved by successively optimizing it along one coordinate direction at a time, i.e., solving much simpler univariate optimization problems in a loop \cite{wright2015coordinate}.
Specifically, for each state $s$, we would need to identify the joint action that can maximize the multivariate function $F_s(a^1,a^2,\ldots,a^N) \equiv Q^i(s,{a^1,a^2,\ldots,a^N};\Theta)$ which, as we have explained above, is shared by all the agents (in the same group).
Using the standard technique of coordinate descent (or more accurately coordinate ascent in our context), the current solution to the $F_s$ optimization problem, $(a^1_t,a^2_t,\ldots,a^N_t)$, can be iteratively improved by finding
\begin{IEEEeqnarray}{rcl}
a^i_{t+1} 
& = & \argmax_{a^i} F_s(a^1_t,\ldots,a^{i-1}_t,a^i,a^{i+1}_t,\ldots,a^N_t) \nonumber \\
& = & \argmax_{a^i} Q^i(s_{t+1},{a^1_t,\ldots,a^{i-1}_t,a^i,a^{i+1}_t,\ldots,a^N_t};\Theta) \nonumber \\
& = & \argmax_{a^i} Q^i(s_{t+1},{a^i,a^{-i}_t};\Theta)  
\end{IEEEeqnarray}
for each variable $a^i$ ($i=1,\dots,N$).
From the perspective of agent $i$, the action to perform at time-step $t+1$, $a^i_{t+1}$, is obtained by fixing the other agents' actions $a^{-i}_t$ and optimizing the objective with respect to its own action $a^i$ only.
Since all the agents simultaneously carry out such coordinate descent updates in parallel at each time-step and the \emph{experience replay} mechanism from DQN \cite{Volodymyr-DQN-Nature-2015} is adopted, the concrete method to estimate the optimal joint action for each state is somewhat similar to the ``asynchronous (parallel) stochastic coordinate descent'' (AsySCD) algorithm which has been proved to have sublinear convergence rate on general convex functions \cite{liu2015asynchronous}, though Q-functions are of course not necessarily convex.
To summarize, the FQL learning process has two kinds of iterative updates interwoven with each other: the temporal difference updates of Q-learning and the asynchronous parallel updates of stochastic coordinate descent.

In other words, for each agent $i$ at non-terminal time-step $t$, the target value $Y_t^i$ is heuristically estimated with the target network ${\widetilde Q}^i$ by keeping all the other agents' actions fixed at their $t$-th time-step:
\begin{equation}
\label{target_value_estimation}
Y_t^i \approx r_t^i + \gamma \cdot \widetilde{Q}^i({s_{t+1}},\argmax_{a^i}{Q^i({s_{t+1}},a^i,a_t^{-i};\Theta)};\widetilde{\Theta}) \ .
\end{equation}

This would significantly reduce the computational complexity from $O \left( \prod_i |\mathcal{A}^i| \right)$ for the combinatorial optimization in Eq.~\ref{target_value_estimation_ideal} to $O \left( \sum_i |\mathcal{A}^i| \right)$ for the simple linear scan of each agent's possible actions in Eq.~\ref{target_value_estimation}.
Although such an aggressive method for action estimation is introduced for efficiency purposes, it turns out to be also empirically very effective for different kinds of MARL tasks, as shown later by our experiments.

\subsection{Computational Complexity}

As we have explained above, all the agents in the same group would share the same factorized Q-function. 
Therefore, a multi-agent system with $G$ groups would only need to maintain $G$ Q-functions, and thus the whole complexity of computation would be merely $G$ times that for one agent's Q-function, no matter how many agents there are in the system.
In practice, $G$ is usually a very small number, and $G=1$ for pure cooperative tasks. 

The factorized Q-function for one agent in a group of size $N$ would behave just like a single-agent DQN \cite{Volodymyr-DQN-Nature-2015}, except that in the former there are $N$ joint actions to be evaluated for the next state at each step of learning while in the latter there is just one single action. 
Nevertheless, using the previously described approximate optimization algorithm, an agent $i$'s best action at the next state $s_{t+1}$ is estimated with the other agents' actions fixed at their current choices $a_t^{-i}$, so at time-step $t$ we could efficiently construct $N$ training examples $\{ Y_t^i,Q^i(s_t,a_t^i,a_t^{-i};\Theta) \}_{i=1}^N$ for the deep neural network. 
Although at each step our proposed multi-agent FQL approach would generate $N$ times more training examples than a single-agent DQN, the total number of training examples required to reach convergence should be quite similar. 
So, the overall computational complexity of training one group of agents in FQL seems to be comparable to that of training a single agent in DQN. 

\section{Experiments}

We evaluate our proposed FQL approach to MARL on two different problems both involving quite a number of agents:
the first is a pure cooperative task while the second is a mixed cooperative-competitive task.

\subsection{The Traffic Game}

\begin{figure*}[htb]
	\centering
	\subfigure[$N=100$]{
		\includegraphics[width=0.48\textwidth]{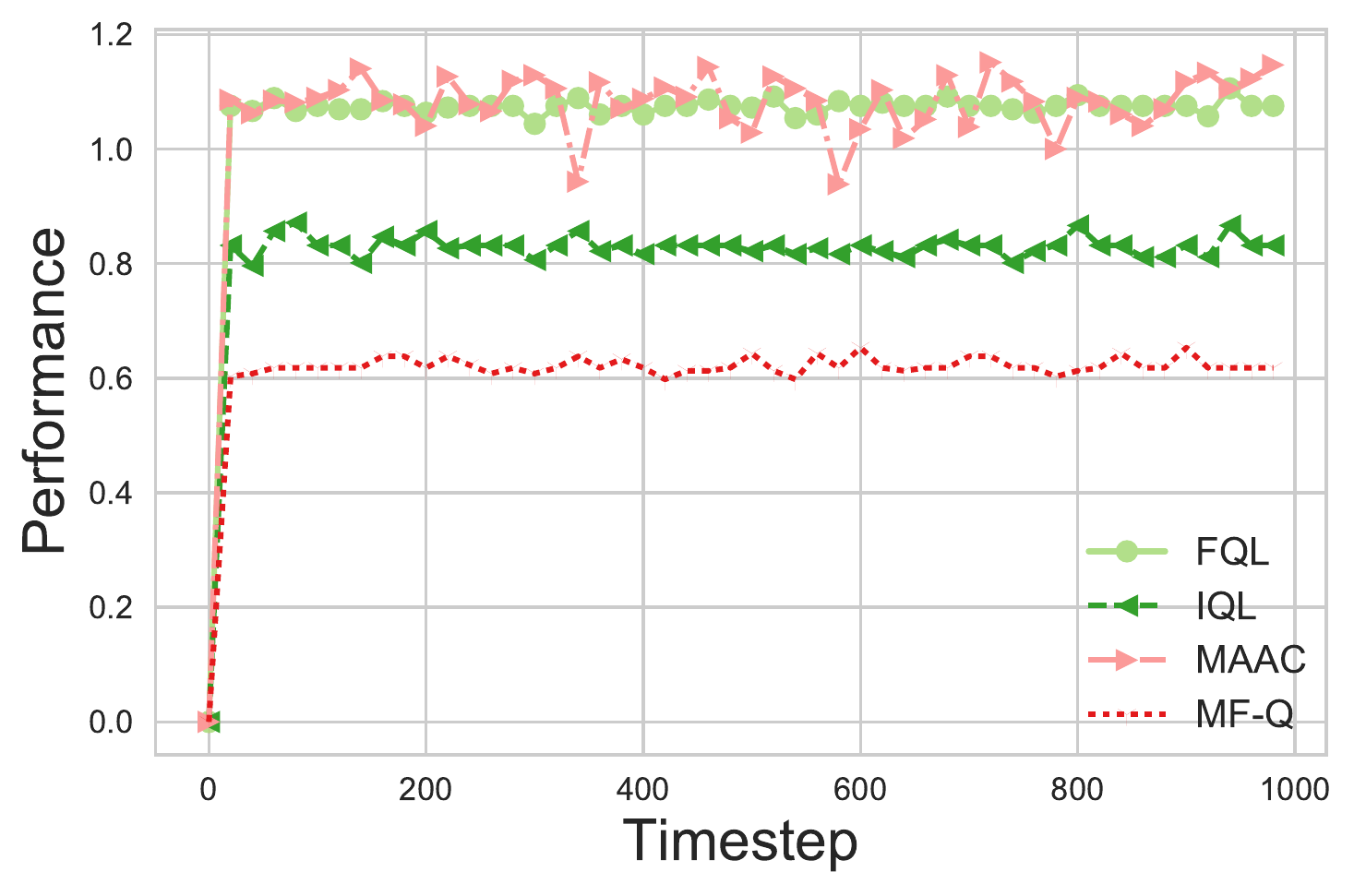}
		\label{fig:GMSD_100_10}}
	\subfigure[$N=500$]{
		\includegraphics[width=0.48\textwidth]{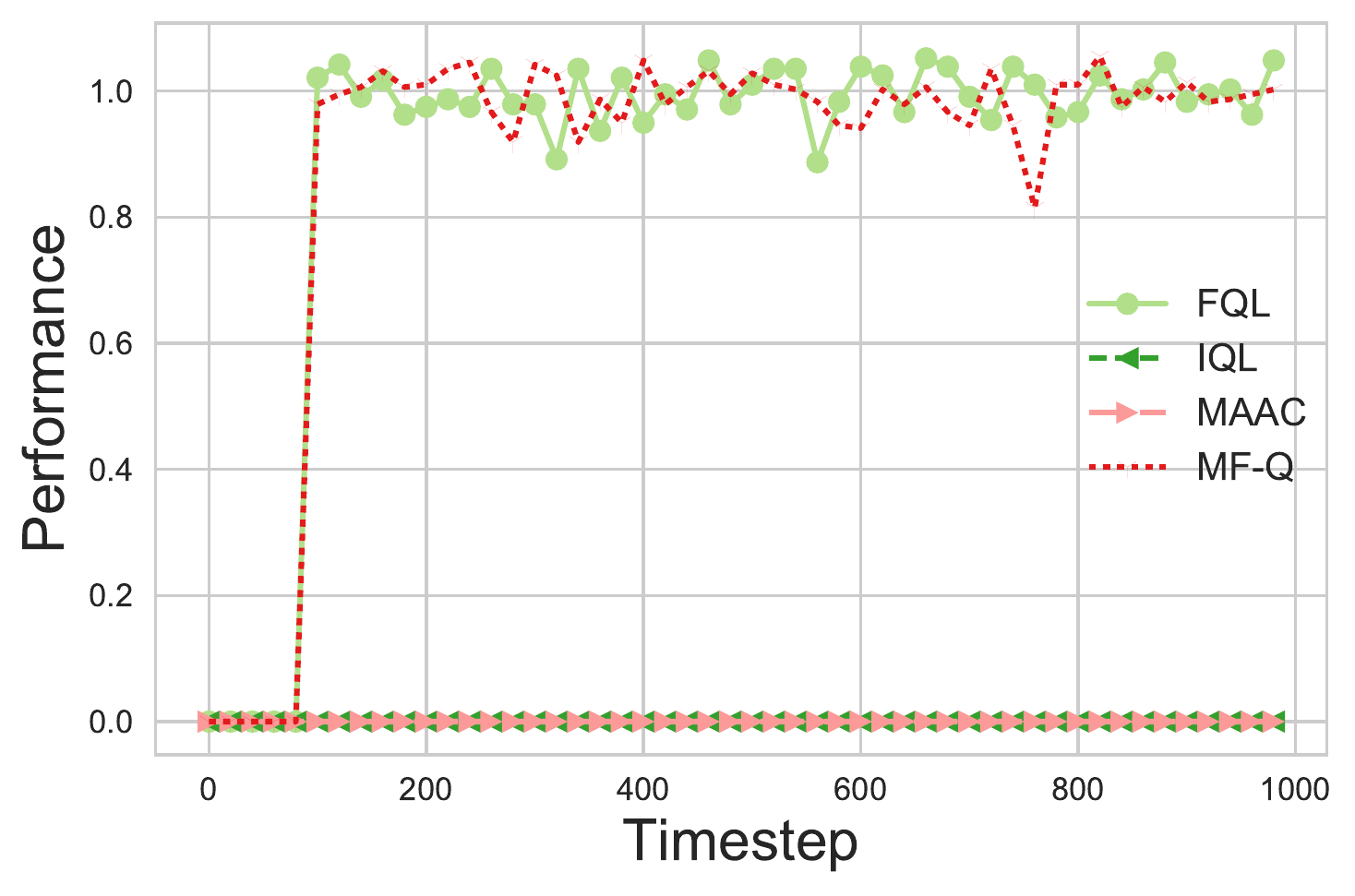}
		\label{fig:GMSD_500_10}}
	\caption{The learning curve of different algorithms (with $N$ agents) in the Traffic Game (Gaussian Squeeze).}
	\label{fig:mgs_exp}
\end{figure*}

\paragraph{Environment.}

Let us consider a resource allocation problem called Gaussian Squeeze \cite{holmesparker2014exploiting} which is inspired by the traffic control task where we want to let as many cars as possible use the available road without causing traffic congestion.
Specifically, $N$ agents need to work together to allocate resources in such a way that the total allocated resources $x = \sum_i {{x_i}}$ is neither too many nor too few, where $x_i$ is the quantity of resources allocated by agent $i$ ($1 \leq i \leq N$).
Given a target quantity of the total allocation $\mu$, we define the reward by a scaled Gaussian function $x \cdot e^{ {- {(x-\mu)}^2}/{\sigma^2} }$ where the parameter $\sigma$ controls the penalty for the deviation of $x$ from the desired value $\mu$.
In the generalized version of this problem, we could have $K$ allocation targets each of which is defined by its own pair of $\mu_k$ and $\sigma_k$ parameters ($1 \leq k \leq K$).
The complete reward function is then given by the sum of the rewards for those $K$ allocation targets $\sum_k x \cdot e^{ {- {(x-\mu_k)}^2}/{\sigma_k^2} } $.

Since in this traffic game the agents do not have state transitions, it is simply a one-state Markov Decision Process aka a Multi-Armed Bandit (MAB) problem.
Nevertheless, it is still very challenging for many agents to cooperate with each other in this game.

In our experiments, each agent can choose an integer value from 0 to 9 as its action (allocation of resources), and the objective consists of two prefixed allocation targets with $(\mu_1=0,\sigma_1=100)$ and $(\mu_2=400,\sigma_2=200)$.

\paragraph{Settings.}

The following popular Q-learning based multi-agent reinforcement learning methods have been used in the experiments to compare with our proposed factorized Q-learning (FQL) approach: independent Q-learning (IQL) \cite{Ming-MARL-ICML-1993,Ardi-MultiagentDQN-PLOSONE-2017}, multi-agent actor-critic (MAAC) \cite{Ryan-MAAC-NIPS-2017}, and mean-field Q-learning (MF-Q) \cite{Yang2018}.
All the competitors employ three-layer perceptrons (feed-forward neural networks) to approximate their Q-functions. 
In particular, our FQL model involves three sub-networks ($Q$, $V$, and $U$) each of which is realized as a three-layer perceptron in exactly the same way. 

\paragraph{Results.}

Fig.~\ref{fig:mgs_exp} shows the experimental results of different algorithms in two scenarios: one with a relatively small number of agents ($N=100)$ and the other with a relatively large number of agents ($N=500$).
As we can see, IQL performed quite well in the former scenario with $100$ agents. 
It is because when $N=100$ agents are independent of each other the sum of their allocations would have the expected value $\mathbb{E}[x] = \sum_{i=1}^N \mathbb{E}[x_i] = N \times \mathbb{E}[x_i] = 100 \times \frac{(0+1+\cdots+9)}{10} = 450$ which happens to be close to one of the allocation targets ($\mu_2=400$). 
However, IQL failed miserably for this cooperative task with $500$ agents. 
These two contrary outcomes confirm that IQL does not have any ability to let multiple agents cooperate with each other. 
With respect to the MF-Q algorithm, its performance is the worst (even inferior to IQL) when there are just $100$ agents, but it becomes almost the best when there are $500$ agents. 
This is reasonable, as the MF-Q algorithm estimates the average effect of actions using the mean-field theory which would be more accurate with more agents according to the law of large numbers. 
The opposite phenomenon is observed for the MAAC algorithm, its performance is one of the best with $100$ agents, but one of the worst (similar to IQL) with $500$ agents. 
This is probably due to the fact that the policy gradients used by MAAC would be harder and harder to be estimated accurately with more and more agents in the system. 
Finally, it is clear that our proposed FQL algorithm is the only one that has achieved the top performance in both scenarios. 
We believe that the advantage of FQL for such a cooperative task can be attributed to its preservation of pairwise interactions among agents in its factorized Q-function formulation.

\subsection{The Battle Game}

\begin{figure}[tb!]
	\centering
	\subfigure[Killing Index]{
		\includegraphics[width=.30\columnwidth]{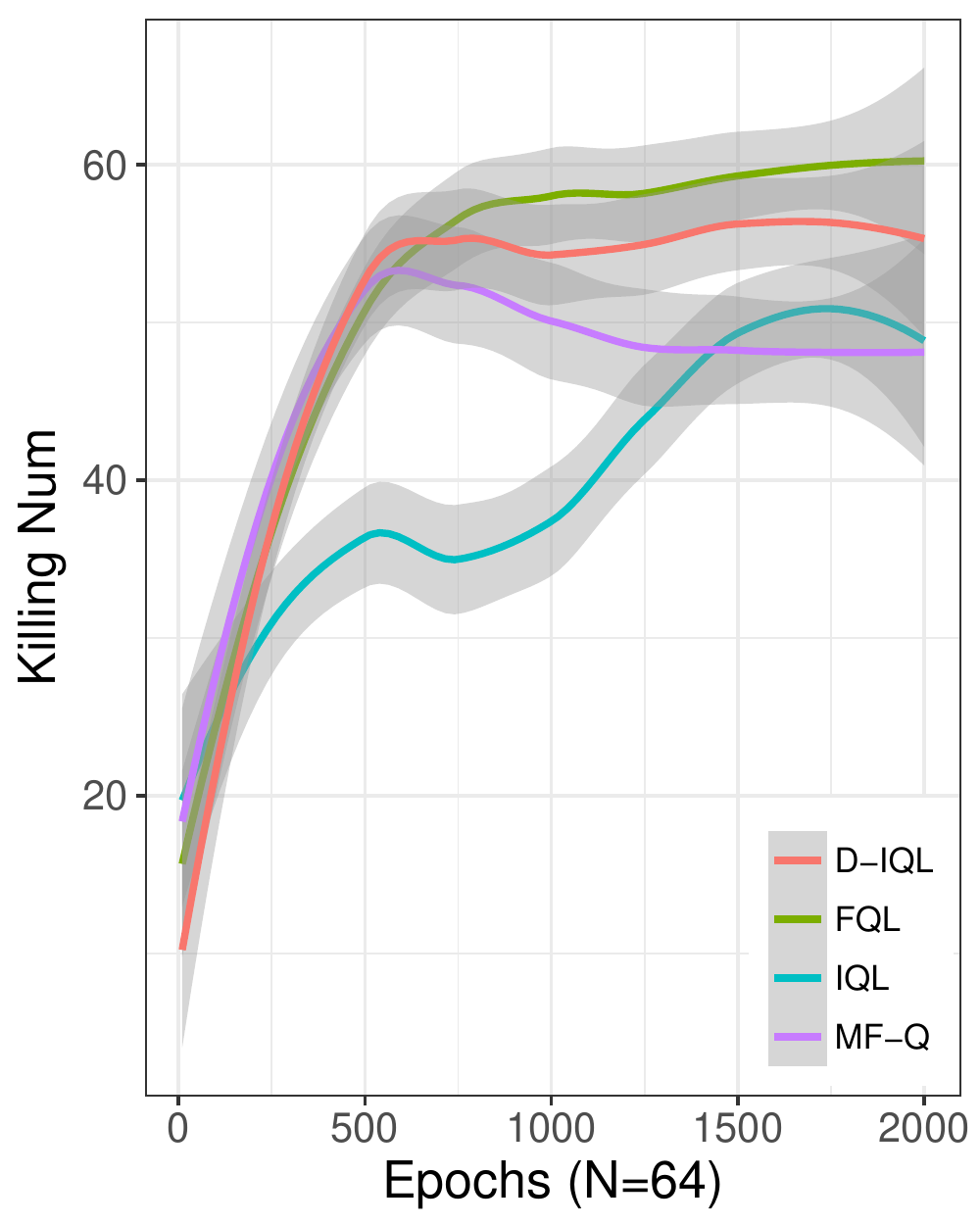}
		\label{k_index_64}}
	\subfigure[Mean Rewards]{
		\includegraphics[width=.30\columnwidth]{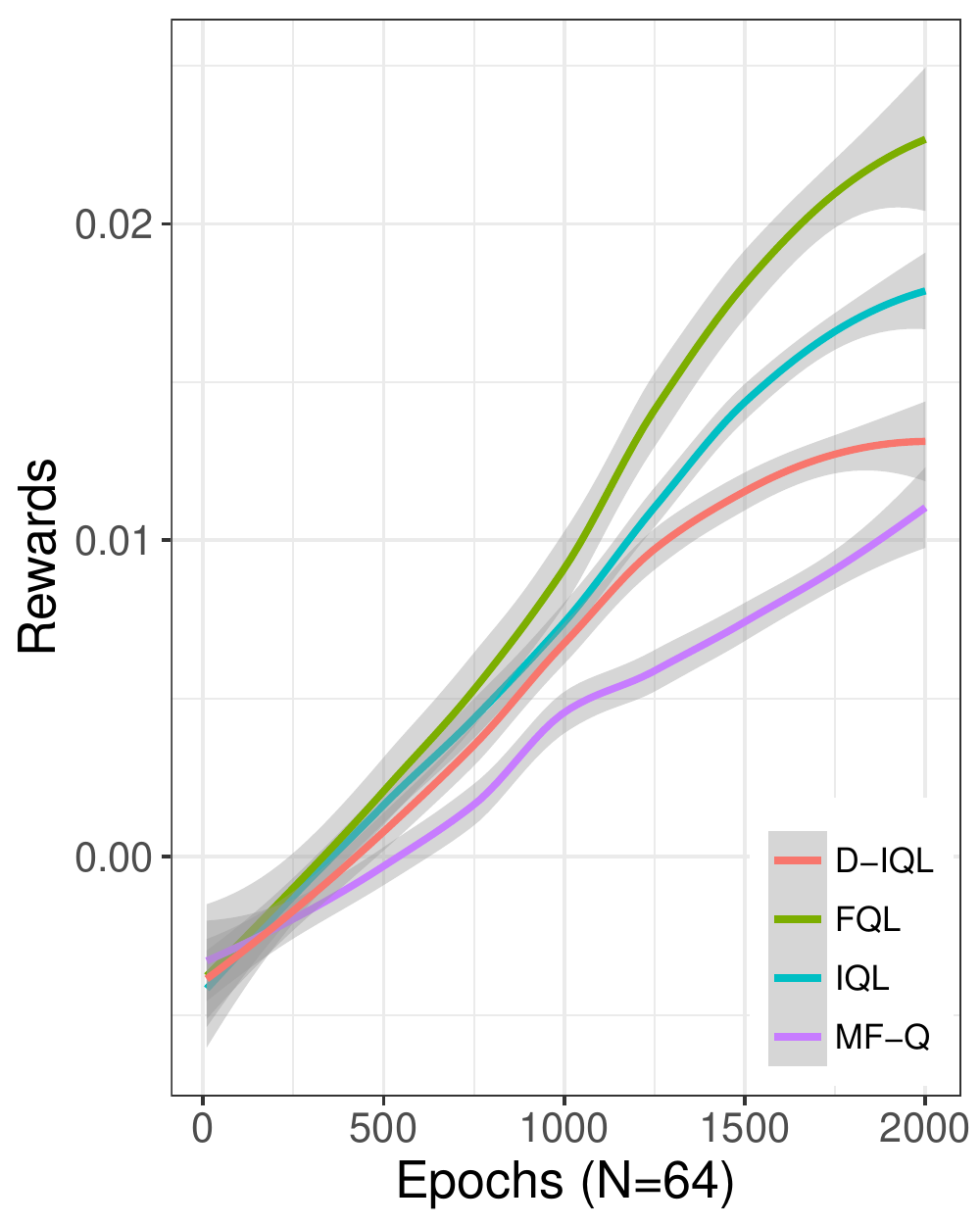}
		\label{mw_index_64}}
	\subfigure[Total Rewards]{
		\includegraphics[width=.30\columnwidth]{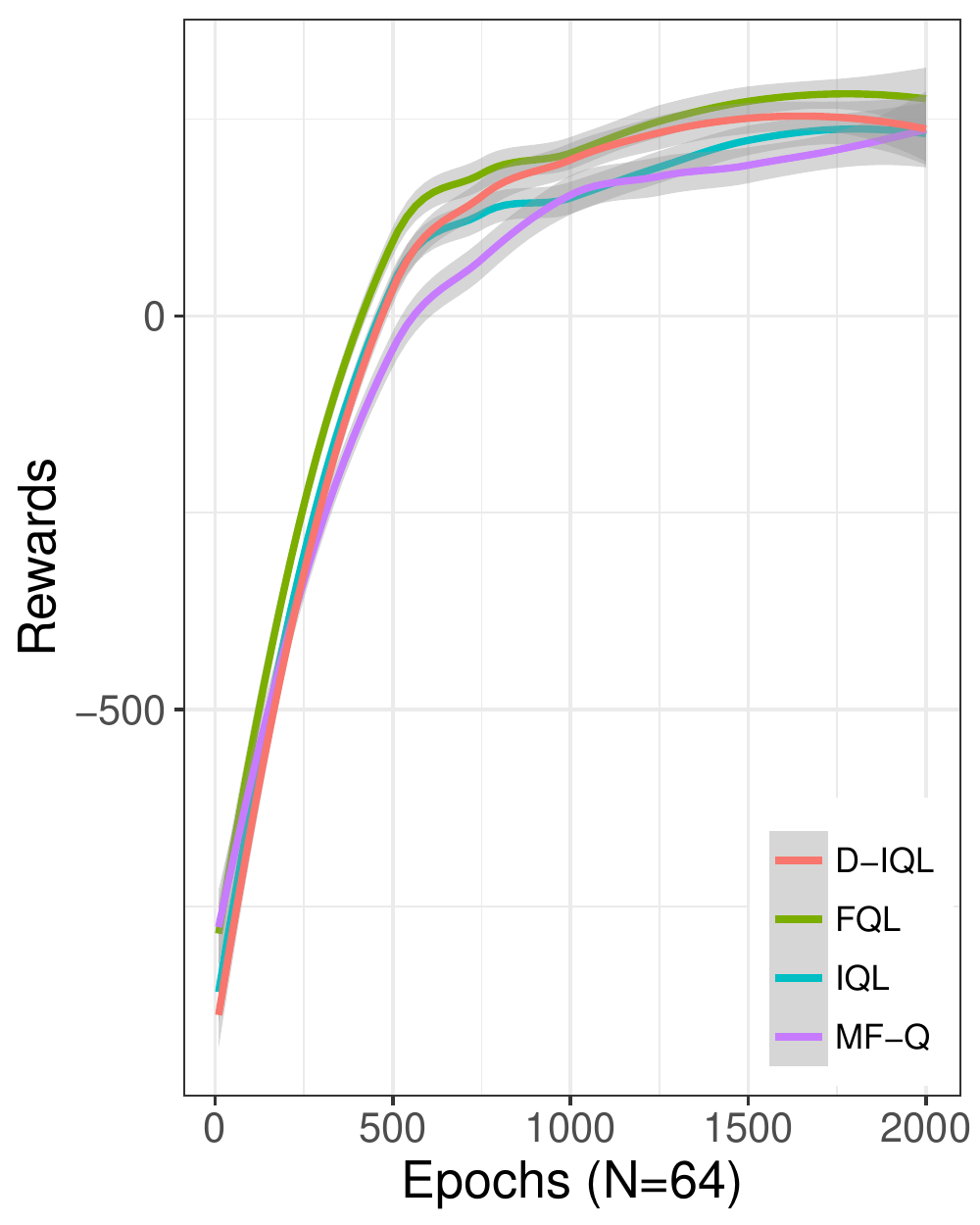}
		\label{tw_index_64}}
	\caption{The learning curve of different algorithms in the Battle Game (recorded every 10 rounds of training with self-play).}
	\label{fig:battle_train}
\end{figure}

\paragraph{Environment.}

The recently emerged open source multi-agent reinforcement learning platform MAgent \cite{Lianmin-MAgent-AAAI-2018} enable us to simulate battles between two armies (groups) in which the soldiers from one army would cooperate with each other to fight against their enemies, i.e., the soldiers from the other army. 
In our experiments, each army consists of $64$ soldiers who would be arrayed in the battlefield (a grid world). 
At each time-step, a soldier would attempt to either move to or attack one of the $8$ neighboring grids. 
The overall objective of an army is to destroy as many enemies as possible. 

\paragraph{Settings.}

The following popular Q-learning based multi-agent reinforcement learning methods have been used in the experiments to compare with our proposed factorized Q-learning (FQL) approach: independent Q-learning (IQL) \cite{Ming-MARL-ICML-1993,Ardi-MultiagentDQN-PLOSONE-2017}, independent Q-learning with the dueling network architecture (D-IQL) \cite{Ziyu-Dueling-Q-Network-ICML-2016}, and mean-field Q-learning (MF-Q) \cite{Yang2018}. 
The MAAC algorithm appeared in the previous game turned out to be incapable of learning to handle a large number of agents in this game (cf. Fig.~\ref{fig:mgs_exp}), therefore we consider D-IQL instead.
The state of each agent consists of the agent's own feature vector which contains its group label, its observation of the grid world, its last reward received, and its last action taken.  
Encoding the group label in the state enables the agent to distinguish friends from foes.
To approximate the Q-functions, all the competitors including our FQL model employ convolutional neural networks (CNN) with the same structure where the local observation is embedded by two convolutional layers plus one fully-connected layer while the feature vector is handled by just one fully-connected layer.
All the models will be trained with 2000 rounds of \emph{self-play}, and then put into one-vs-one battles against each other.
The full description of the implementation details are provided in the supplementary material.


\begin{figure}[tb!]
	\centering
	\subfigure[Mean Rewards]{
		\includegraphics[width=0.47\columnwidth]{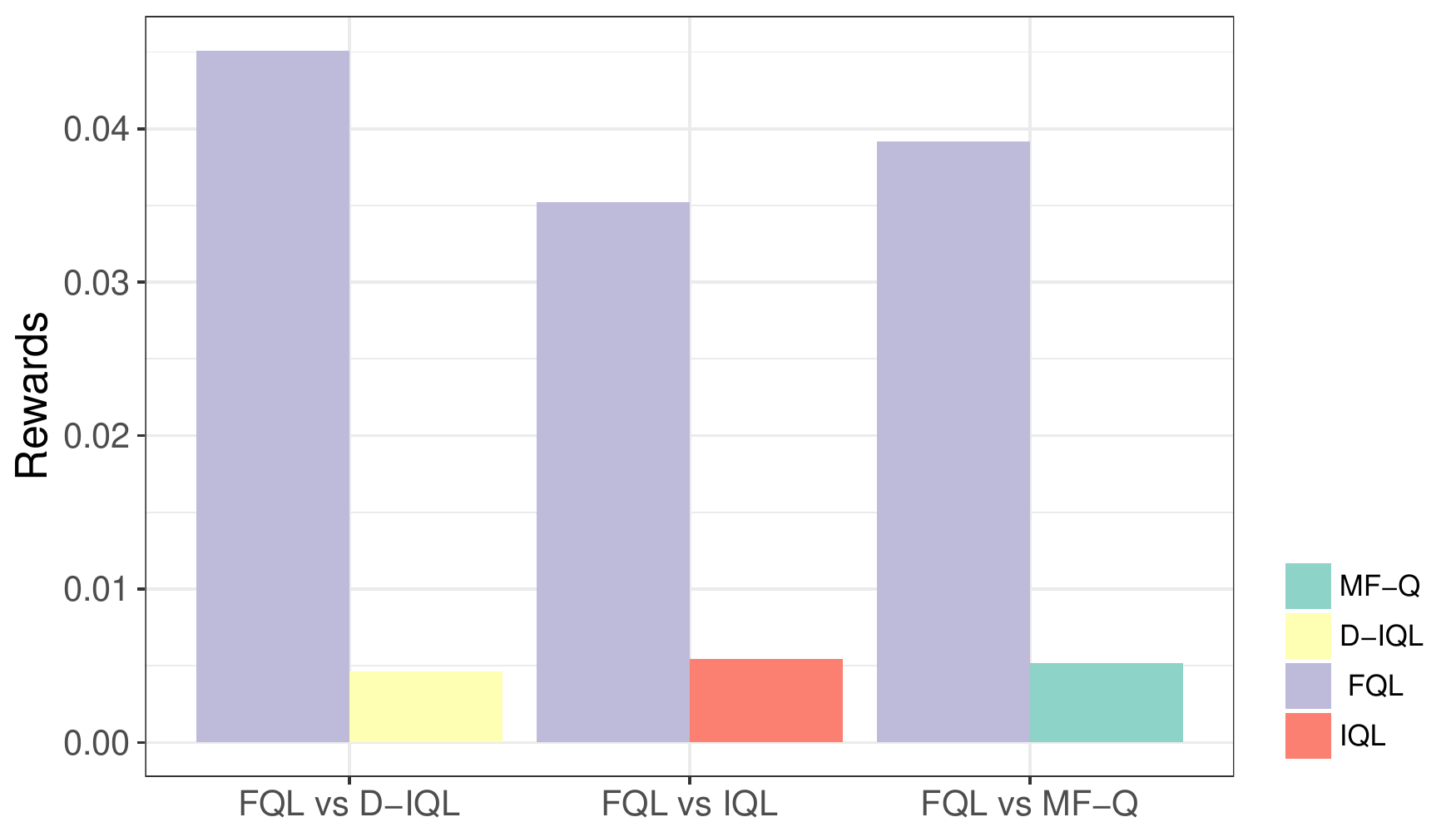}\label{compare_mw_index_64}}
	\subfigure[Total Rewards]{
		\includegraphics[width=0.47\columnwidth]{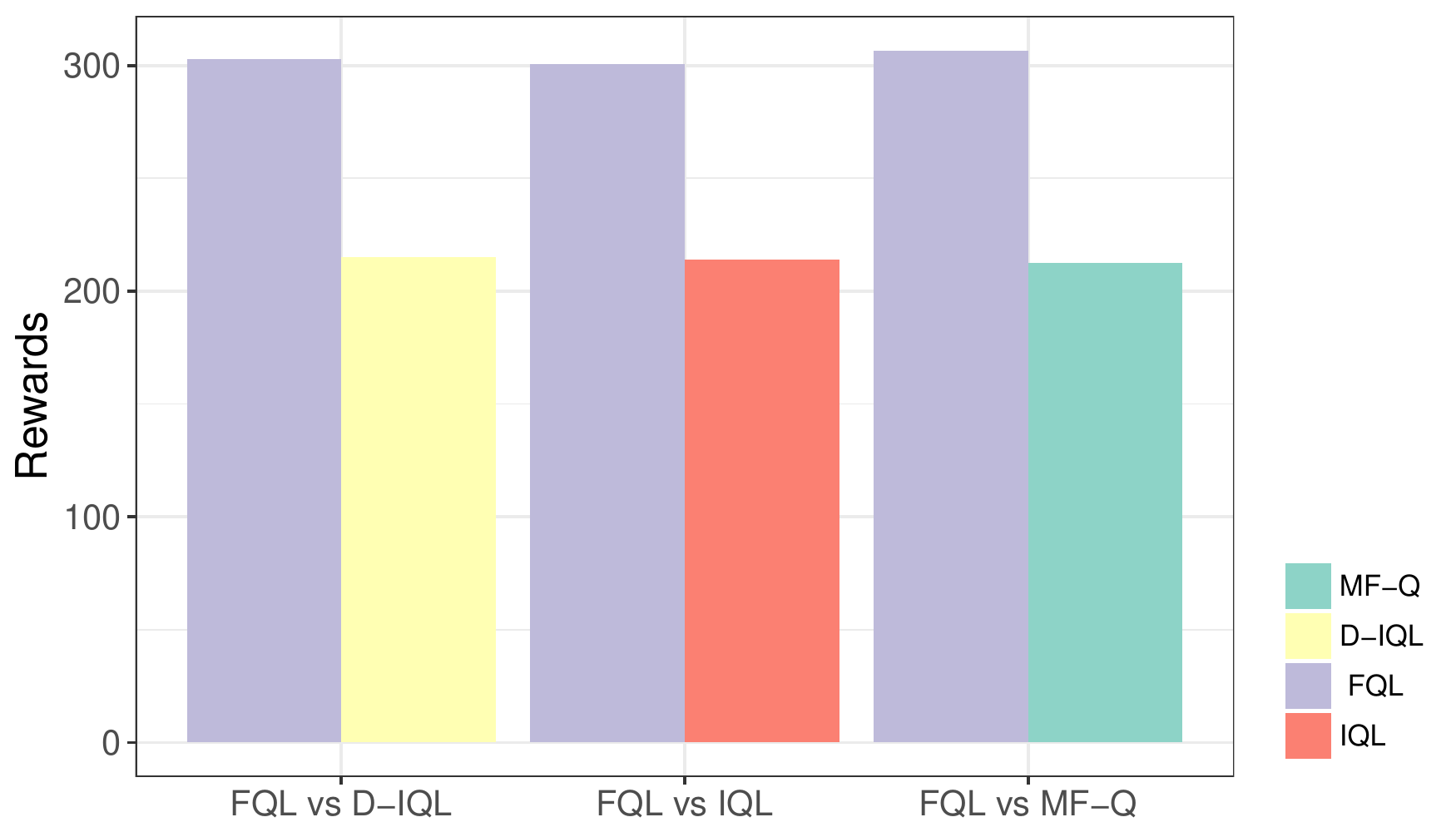}
		\label{compare_tw_index_64}}
	\caption{The performance of FQL competing against each baseline algorithm in the Battle Game (when each agent considers \emph{all~the~other} agents). The reported results are the average values over 100 battles for each comparative experiment.}
	\label{fig:battle_compare}
 \end{figure}
 \begin{figure}[tb!]
	\centering
	\subfigure[Mean Rewards]{
		\includegraphics[width=0.47\columnwidth]{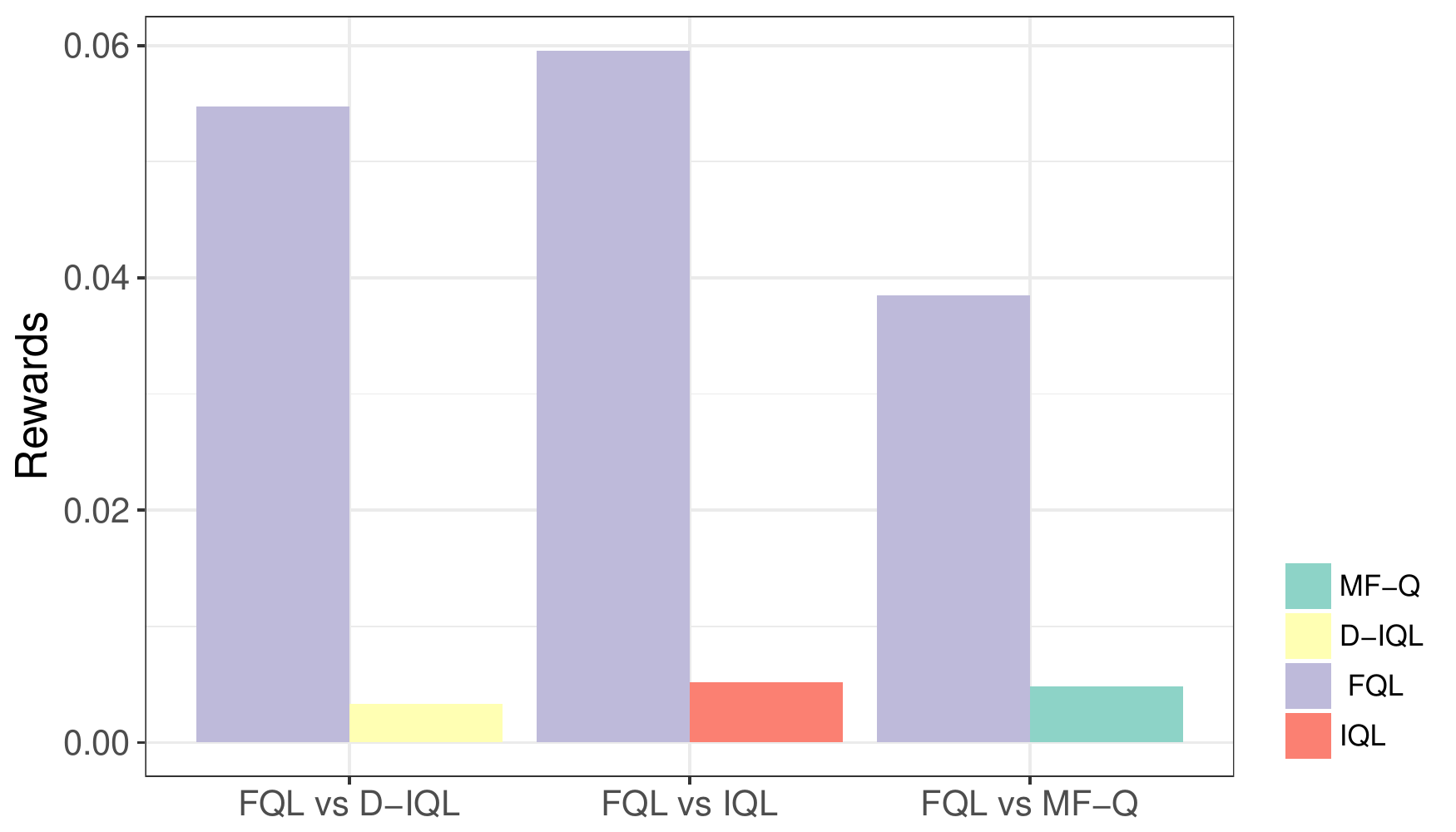}
		\label{compare_mw_index_64_local}}
	\subfigure[Total Rewards]{
		\includegraphics[width=0.47\columnwidth]{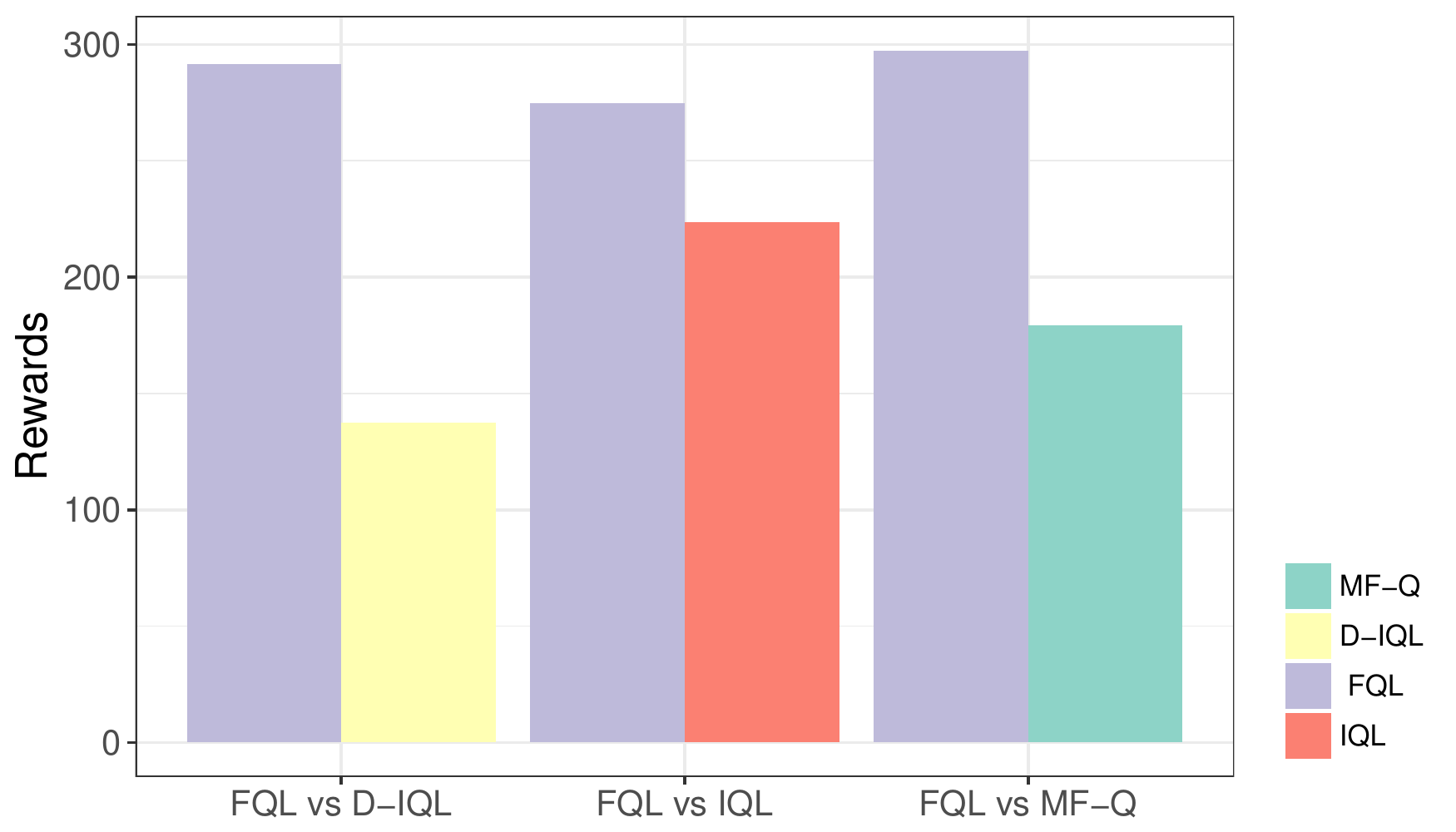}
		\label{compare_tw_index_64_local}}
	\caption{The performance of FQL competing against each baseline algorithm in the Battle Game (when each agent considers \emph{the~neighboring} agents). The reported results are the average values over 100 battles for each comparative experiment.}
	\label{fig:battle_compare_local}
\end{figure}

\paragraph{Results.}

Fig.~\ref{fig:battle_train} shows the learning curves of different algorithms w.r.t. three different performance metrics: the number of enemies killed in the battle (``Killing Index''), the average reward obtained by each soldier (``Mean Rewards''), and the total reward for the entire army (``Total Rewards'').
These three different metrics reflect different aspects of this mixed cooperative-competitive game.
Specifically,
the ``Killing Index'' indicates how fierce the battle was;
the ``Mean Rewards'' defined as $R_\text{mean}=\frac{1}{N}\sum_{i=1}^{N}(R_i/T_i)$ where $T_i$ is the survival time of agent $i$ and $R_i$ is the total reward during agent $i$'s survival time, represents how strong on average an individual soldier was; and 
the ``Total Rewards'' shows how effective the teamwork of the army was.
It is clear that in the policy learning stage, the FQL model could be trained more quickly and also reach a higher capacity than the other models (i.e., IQL, D-IQL, and MF-Q) in terms of all the above mentioned metrics.

Fig.~\ref{fig:battle_compare} further shows the cross-comparison experimental results between FQL and the three competitors in the policy execution stage (averaged over $100$ one-vs-one battles). 
The competitive advantage of FQL over the other models can be seen clearly. 
It suggests that the factorized Q-function could indeed capture the most important interactions among agents and thus encourage cooperation within the group.

When the group size $N$ becomes larger, the training of the FQL model will become more computationally expensive, because each agent would need to know not only the current state but also the last actions of all the other $N-1$ agents in the same group for each step of Q-function update.
Therefore, we go further to investigate what happens if each agent can only remember the last actions of the \emph{neighboring} agents (i.e., those within a radius of $13$).
In such a neighborhood-level decentralized paradigm, the learning and execution of the FQL model would be a lot more efficient than in the centralized paradigm.
The experimental results in Fig.~\ref{fig:battle_compare_local} demonstrate that this partially decentralized version of FQL could still achieve pretty good results in comparison with the other models. 
It suggests that FQL has the potential to scale up to even larger multi-agent systems.

\section{Conclusions}

The main contribution of this paper is a novel factorized formulation of the joint state-action Q-function which makes reinforcement learning with many agents computationally feasible.
The experimental results suggest that although our proposed FQL model relies on several aggressive simplifications to ensure the efficiency, it is surprisingly effective as shown by its performance for both a pure cooperative task and a mixed cooperative-competitive task.
An open research question is whether the FQL algorithm is guaranteed to converge. 
The answer seems to be ``yes'' based on the empirical evidence that FQL has always converged in our experiments, but the theoretical proof is left for future work.

\bibliographystyle{named}
\bibliography{dai_2019}

\begin{thebibliography}{}

\bibitem[\protect\citeauthoryear{Blume}{1993}]{Lawrence-Strategic-Interaction-Games-1993}
Lawrence~E. Blume.
\newblock The statistical mechanics of strategic interaction.
\newblock {\em Games and Economic Behavior}, pages 387--424, 1993.

\bibitem[\protect\citeauthoryear{Claus and Boutilier}{1998}]{Claus1998}
Caroline Claus and Craig Boutilier.
\newblock The dynamics of reinforcement learning in cooperative multiagent
  systems.
\newblock In {\em Proceedings of the 15th National Conference on Artificial
  Intelligence (AAAI) and 10th Innovative Applications of Artificial
  Intelligence Conference (IAAI)}, pages 746--752, Madison, WI, US, 1998.

\bibitem[\protect\citeauthoryear{Foerster \bgroup \em et al.\egroup
  }{2016}]{Jakob-RIAL-NIPS-2016}
Jakob~N. Foerster, Yannis~M. Assael, Nando de~Freitas, and Shimon Whiteson.
\newblock Learning to communicate with deep multi-agent reinforcement learning.
\newblock In {\em NIPS}, pages 2137--2145, 2016.

\bibitem[\protect\citeauthoryear{HolmesParker \bgroup \em et al.\egroup
  }{2014}]{holmesparker2014exploiting}
Chris HolmesParker, Matthew~E. Taylor, Yusen Zhan, and Kagan Tumer.
\newblock Exploiting structure and agent-centric rewards to promote
  coordination in large multiagent systems.
\newblock In {\em Adaptive and Learning Agents Workshop (ALA)}, 2014.

\bibitem[\protect\citeauthoryear{Hu and
  Wellman}{2003}]{Junling-NashQLearning-JMLR-2003}
Junling Hu and Michael~P. Wellman.
\newblock Nash {Q}-learning for general-sum stochastic games.
\newblock {\em Journal of Machine Learning Research (JMLR)}, 4:1039--1069,
  2003.

\bibitem[\protect\citeauthoryear{Kok and
  Vlassis}{2004}]{Jelle-Sparse-Cooperative-Q-ICML-2004}
Jelle~R. Kok and Nikos~A. Vlassis.
\newblock Sparse cooperative {Q}-learning.
\newblock In {\em ICML}, 2004.

\bibitem[\protect\citeauthoryear{Lample and
  Chaplot}{2017}]{Guillaume-DRQN-FPS-AAAI-2017}
Guillaume Lample and Devendra~Singh Chaplot.
\newblock Playing {FPS} games with deep reinforcement learning.
\newblock In {\em AAAI}, pages 2140--2146, 2017.

\bibitem[\protect\citeauthoryear{Littman}{1994}]{Littman-MarkovGames-ICML-1994}
Michael~L. Littman.
\newblock {Markov} games as a framework for multi-agent reinforcement learning.
\newblock In {\em ICML}, pages 157--163, 1994.

\bibitem[\protect\citeauthoryear{Littman}{2001a}]{Michael-FFQ-ICML-2001}
Michael~L. Littman.
\newblock {Friend-or-Foe} {Q}-learning in general-sum games.
\newblock In {\em ICML}, pages 322--328, 2001.

\bibitem[\protect\citeauthoryear{Littman}{2001b}]{Michael-Team-Q-CognitiveSystems-2001}
Michael~L. Littman.
\newblock Value-function reinforcement learning in {Markov} games.
\newblock {\em Cognitive Systems Research}, pages 55--66, 2001.

\bibitem[\protect\citeauthoryear{Liu \bgroup \em et al.\egroup
  }{2015}]{liu2015asynchronous}
Ji~Liu, Stephen~J Wright, Christopher R{\'e}, Victor Bittorf, and Srikrishna
  Sridhar.
\newblock An asynchronous parallel stochastic coordinate descent algorithm.
\newblock {\em The Journal of Machine Learning Research (JMLR)},
  16(1):285--322, 2015.

\bibitem[\protect\citeauthoryear{Lowe \bgroup \em et al.\egroup
  }{2017}]{Ryan-MAAC-NIPS-2017}
Ryan Lowe, Yi~Wu, Aviv Tamar, Jean Harb, Pieter Abbeel, and Igor Mordatch.
\newblock Multi-agent actor-critic for mixed cooperative-competitive
  environments.
\newblock In {\em NIPS}, pages 6382--6393, 2017.

\bibitem[\protect\citeauthoryear{Melo}{2001}]{Francisco-Convergence-Report-2001}
Francisco~S. Melo.
\newblock Convergence of {Q}-learning: A simple proof.
\newblock {\em Tech. Rep.}, pages 1--4, 2001.

\bibitem[\protect\citeauthoryear{Mikolov \bgroup \em et al.\egroup
  }{2013}]{mikolov2013efficient}
Tomas Mikolov, Kai Chen, Greg Corrado, and Jeffrey Dean.
\newblock Efficient estimation of word representations in vector space.
\newblock {\em arXiv preprint arXiv:1301.3781}, 2013.

\bibitem[\protect\citeauthoryear{Mnih \bgroup \em et al.\egroup
  }{2015}]{Volodymyr-DQN-Nature-2015}
Volodymyr Mnih, Koray Kavukcuoglu, David Silver, Andrei~A. Rusu, Joel Veness,
  Marc~G. Bellemare, Alex Graves, Martin Riedmiller, Andreas~K. Fidjeland,
  Georg Ostrovski, Stig Petersen, Charles Beattie, Amir Sadik, Ioannis
  Antonoglou, Helen King, Dharshan Kumaran, Daan Wierstra, Shane Legg, and
  Demis Hassabis.
\newblock Human-level control through deep reinforcement learning.
\newblock {\em Nature}, 518(7540):529--533, 2015.

\bibitem[\protect\citeauthoryear{Ong}{2015}]{ong2015value}
Hao~Yi Ong.
\newblock Value function approximation via low-rank models.
\newblock {\em arXiv preprint arXiv:1509.00061}, 2015.

\bibitem[\protect\citeauthoryear{Rashid \bgroup \em et al.\egroup
  }{2018}]{Tabish-QMIX-CoRR-2018}
Tabish Rashid, Mikayel Samvelyan, Christian~Schroder de~Witt, Gregory Farquhar,
  Jakob~N. Foerster, and Shimon Whiteson.
\newblock {QMIX}: Monotonic value function factorisation for deep multi-agent
  reinforcement learning.
\newblock {\em arXiv preprint arXiv:1803.11485}, 2018.

\bibitem[\protect\citeauthoryear{Rendle and
  Schmidt-Thieme}{2010}]{Steffen-PITensor-WSDM-2010}
Steffen Rendle and Lars Schmidt-Thieme.
\newblock Pairwise interaction tensor factorization for personalized tag
  recommendation.
\newblock {\em WSDM}, pages 81--90, 2010.

\bibitem[\protect\citeauthoryear{Rendle}{2012}]{Steffen-FM-TIST-2012}
Steffen Rendle.
\newblock Factorization machines with {libFM}.
\newblock {\em ACM Transactions on Intelligent Systems and Technology (TIST)},
  pages 1--22, 2012.

\bibitem[\protect\citeauthoryear{Sallans and
  Hinton}{2004}]{sallans2004reinforcement}
Brian Sallans and Geoffrey~E Hinton.
\newblock Reinforcement learning with factored states and actions.
\newblock {\em Journal of Machine Learning Research (JMLR)}, 5(Aug):1063--1088,
  2004.

\bibitem[\protect\citeauthoryear{Sunehag \bgroup \em et al.\egroup
  }{2017}]{Peter-VDNs-CoRR-2017}
Peter Sunehag, Guy Lever, Audrunas Gruslys, Wojciech~Marian Czarnecki,
  Vinicius~Flores Zambaldi, Max Jaderberg, Marc Lanctot, Nicolas Sonnerat,
  Joel~Z. Leibo, Karl Tuyls, and Thore Graepel.
\newblock Value-decomposition networks for cooperative multi-agent learning.
\newblock {\em arXiv preprint arXiv:1706.05296}, 2017.

\bibitem[\protect\citeauthoryear{Sutton and
  Barto}{1998}]{sutton1998reinforcement}
Richard~S Sutton and Andrew~G Barto.
\newblock {\em Reinforcement Learning: An Introduction}.
\newblock MIT Press, 1998.

\bibitem[\protect\citeauthoryear{Tampuu \bgroup \em et al.\egroup
  }{2017}]{Ardi-MultiagentDQN-PLOSONE-2017}
Ardi Tampuu, Tambet Matiisen, Dorian Kodelja, Ilya Kuzovkin, Kristjan Korjus,
  Juhan Aru, Jaan Aru, and Raul Vicente.
\newblock Multiagent cooperation and competition with deep reinforcement
  learning.
\newblock {\em PLoS ONE}, 2017.

\bibitem[\protect\citeauthoryear{Tan}{1993}]{Ming-MARL-ICML-1993}
Ming Tan.
\newblock Multi-agent reinforcement learning: Independent versus cooperative
  agents.
\newblock In {\em ICML}, pages 330--337, 1993.

\bibitem[\protect\citeauthoryear{Tesauro}{2003}]{Gerald-Hyper-Q-NIPS-2003}
Gerald Tesauro.
\newblock Extending {Q}-learning to general adaptive multi-agent systems.
\newblock In {\em NIPS}, pages 871--878, 2003.

\bibitem[\protect\citeauthoryear{Uther and Veloso}{1997}]{Uther1997}
William Uther and Manuela Veloso.
\newblock Adversarial reinforcement learning.
\newblock Technical report, CMU-CS-03-107, 1997.

\bibitem[\protect\citeauthoryear{van Hasselt \bgroup \em et al.\egroup
  }{2016}]{Hado-DDQN-AAAI-2016}
Hado van Hasselt, Arthur Guez, and David Silver.
\newblock Deep reinforcement learning with double {Q}-learning.
\newblock In {\em AAAI}, pages 2094--2100, 2016.

\bibitem[\protect\citeauthoryear{van
  Hasselt}{2010}]{Hado-DoubleQLearning-NIPS-2010}
Hado van Hasselt.
\newblock Double {Q}-learning.
\newblock In {\em NIPS}, pages 2613--2621, 2010.

\bibitem[\protect\citeauthoryear{Wang \bgroup \em et al.\egroup
  }{2016}]{Ziyu-Dueling-Q-Network-ICML-2016}
Ziyu Wang, Tom Schaul, Matteo Hessel, Hado van Hasselt, Marc Lanctot, and Nando
  de~Freitas.
\newblock Dueling network architectures for deep reinforcement learning.
\newblock In {\em ICML}, pages 1995--2003, 2016.

\bibitem[\protect\citeauthoryear{Watkins and
  Dayan}{1992}]{Christopher-Technical-QLearning-ML-1992}
Christopher J. C.~H. Watkins and Peter Dayan.
\newblock Technical note: {Q}-learning.
\newblock {\em Machine Learning}, 8:279--292, 1992.

\bibitem[\protect\citeauthoryear{Watkins}{1989}]{Christopher-PhD-Thesis-1989}
Christopher J. C.~H. Watkins.
\newblock Learning from delayed rewards.
\newblock {\em PhD Thesis, University of Cambridge, England}, 1989.

\bibitem[\protect\citeauthoryear{Wright}{2015}]{wright2015coordinate}
Stephen~J Wright.
\newblock Coordinate descent algorithms.
\newblock {\em Mathematical Programming}, 151(1):3--34, 2015.

\bibitem[\protect\citeauthoryear{Yang \bgroup \em et al.\egroup
  }{2018}]{Yang2018}
Yaodong Yang, Rui Luo, Minne Li, Ming Zhou, Weinan Zhang, and Jun Wang.
\newblock Mean field multi-agent reinforcement learning.
\newblock In {\em ICML}, Stockholm, Sweden, 2018.

\bibitem[\protect\citeauthoryear{Zheng \bgroup \em et al.\egroup
  }{2018}]{Lianmin-MAgent-AAAI-2018}
Lianmin Zheng, Jiacheng Yang, Han Cai, Weinan Zhang, Jun Wang, and Yong Yu.
\newblock {MAgent}: {A} many-agent reinforcement learning platform for
  artificial collective intelligence.
\newblock In {\em AAAI}, 2018.

\end{thebibliography}

\end{document}